\documentclass{ifacconf}

\usepackage{graphicx}      % include this line if your document contains figures
\usepackage{natbib}        % required for bibliography
\usepackage{mathrsfs,amssymb,amsmath}

\newcommand\diag[1]{\operatorname{diag}({#1})}

\newcommand{\unitvector}{e}

\newcommand{\tr}{\operatorname{tr}}

\begin{document}
\begin{frontmatter}

\title{Exploring the Limits of Open Quantum Dynamics II: Gibbs-Preserving Maps from the Perspective of Majorization\thanksref{footnoteinfo}} 

\thanks[footnoteinfo]{The project was supported i.a.\ by Excellence Network of Bavaria under ExQM and is
part of {\em Munich Quantum Valley} of the Bavarian State Government with funds from Hightech Agenda {\em Bayern Plus}.}

\author[First]{Frederik vom Ende} 
%\author[Second]{Second B. Author, Jr.} 
%\author[Third]{Third C. Author}

\address[First]{Dept.~Chem., Lichtenbergstra{\ss}e 4, 85747 Garching, Germany  \&\\ Munich Centre for Quantum Science and Technology (MCQST),  Schellingstra{\ss}e~4, 80799~M{\"u}nchen, Germany\\  (e-mail: frederik.vom-ende@tum.de).}
%\address[Second]{Colorado State University, 
%   Fort Collins, CO 80523 USA (e-mail: author@lamar. colostate.edu)}
%\address[Third]{Electrical Engineering Department, 
%   Seoul National University, Seoul, Korea, (e-mail: author@snu.ac.kr)}

\date={date: \today}

\begin{abstract}                % Abstract of not more than 250 words.
Motivated by reachability questions in coherently controlled open quantum systems coupled to a thermal bath%in a switchable manner
, as well as recent progress in the field of thermo-/vector-majorization % [\cite{vomEnde19polytope}]
we generalize classical majorization from unital quantum channels to channels with an arbitrary fixed point $D$ of full rank.
Such channels preserve some Gibbs-state and thus play an important role in the resource theory of quantum thermodynamics, in particular in thermo-majorization.

$\qquad$Based on this we investigate $D$-majorization on matrices in terms of its topological and order properties, such as existence of unique maximal and minimal elements, etc. Moreover we characterize $D$-majorization in the qubit case via the trace norm and elaborate on why this is a challenging task when going beyond two dimensions.
\end{abstract}


\end{frontmatter}

%===============================================================================
%
\section{Introduction}

Studying reachable sets of control systems is necessary to ensure
well-posedness of a large class of (optimal) control tasks. In
\cite{CDC19} toy models on the standard simplex of probability vectors
were studied in order to answer reachability questions of controlled $n$-level systems coupled to a bath of finite temperature such that the coupling can be switched on and off.
  % where said bath can be switched on and off %acted as an additional control
If the closed (unitary) part of the system can be fully controlled and the bath has temperature $T=0$ then every quantum state\footnote{A quantum state is a positive semi-definite matrix of unit trace.} can be reached approximately from every initial state (that is, perhaps not exactly but at least with arbitrary precision). For $T=\infty$ an upper bound can be obtained by classical majorization techniques. For more details on this we refer to the first part of this talk: \textit{Exploring the Limits of Open Quantum Dynamics I: Motivation, First Results from Toy Models to Applications}, as well as Section \ref{sec_reach}.

%\marginpar{den Einschub kann man auch komplett weglassen!}
An obvious follow-up question is what can be said---if one can say anything at all---about the reachable set of such a system for $0<T<\infty$? Even within the
simplified diagonal toy model (cf.~Part I) this is a rather difficult task and it seems that the notion necessary to handle such problems requires a more
general form of majorization:
\section{On the Road to $D$-Majorization}
\subsection{$d$-Majorization on Vectors}\label{sec:d_maj_vec}
%\section{Generalizing $d$-Majorization to Square Matrices}\label{sec:matrix_d_maj}

Majorization relative to a strictly positive vector $d\in\mathbb R_{++}^n$, as introduced by \cite{Veinott71} 
and in the quantum regime by \cite{Ruch78} is defined as follows: a vector
$y$ is said to $d$-majorize $x$, denoted by $x\prec_d y$, if there exists a $d$-stochastic
% \footnote{A matrix $A\in\mathbb R^{n\times n}$ is said to be \textit{column stochastic} if all its entries are non-negative and 
% $\sum\nolimits_{i=1}^n A_{ij}=1$ for all $j=1,\ldots,n$, so the entries of each column sum up to one.}
  matrix $A\in\mathbb R^{n\times n}$ with $x=Ay$. Recall that $A\in\mathbb R^{n\times n}$ is $d$-stochastic if all its entries
  are non-negative and $Ad = d$, $e^TA = e^T$ with $e^T := (1,\ldots,1)^T$.
%\marginpar{Dies wuerde die Fussnote vermeiden}
A variety of characterizations of $\prec_d$ and $d$-stochastic matrices can be found 
in the work of \cite{Joe90} or \cite{vomEnde19polytope}% or Proposition \ref{lemma_char_d_vec} below
. The most useful for numerical purposes is the following: $x\prec_d y$ if and only if $\sum_{j=1}^nx_j=\sum_{j=1}^ny_j$ and $\|d_ix-y_id\|_1\leq\|d_iy-y_id\|_1$ for all $i=1,\ldots,n $, where $\|z\|_1=\sum_{j=1}^n |z_j|$ is the usual vector-1-norm.

Classical majorization $\prec$, that is, $x\prec y$ for $x,y\in\mathbb R^n$, is originally defined via ordering $x,y$ decreasingly and then comparing partial sums:
%\begin{equation}\label{eq:partial_sum}
$
\sum\nolimits_{j=1}^kx_j^\cdot\leq\sum\nolimits_{j=1}^ky_j^\cdot
$
%\end{equation}
for all $k=1,\ldots,n-1$ as well as $\sum_{j=1}^nx_j=\sum_{j=1}^ny_j$.
%equality in \eqref{eq:partial_sum} for $k=n$
For more on %the topic of 
vector majorization we refer to Ch.~1 \& 2 of \cite{MarshallOlkin}. In particular it is well-known that setting $d=e$ in the definition of $d$-majorization recovers $\prec$ ---which also shows that the definition via partial sums cannot extend beyond $e^T$: as soon as two entries in $d\in\mathbb R_{++}^n$ differ one loses permutation invariance and reordering the vectors $x,y$ makes a conceptual difference.

%The same ``arrangement'' problem occurs when generalizing classical vector majorization---\textcolor{blue}{fr\"uher und mit Formel which was originally defined via comparing the partial sums of the decreasingly rearranged vectors $x,y$}---to arbitrary weight vectors $d$. This led to the unambiguous definition (or rather characterization) via $d$-stochastic matrices from Section \ref{sec:d_maj_vec}.

The above $1$-norm characterization allows to rewrite the $d$-majorization polytope $M_d(y):=\{x\in\mathbb R^n\,|\,x\prec_d y\}$ for any $y\in\mathbb R^n$ as the set of solutions to a nicely structured vector inequality $\mathcal Mx\leq b(y)$, $\mathcal M\in\mathbb R^{2^n\times n}$. This %$\mathscr H$-
description of $d$-majorization enables a proof of the existence of an extremal point $z\in M_d(y)$ such that $M_d(y)\subseteq M_e(z)$, i.e.~there exists some $z\prec_d y$ which classically majorizes all $x\in M_d(y)$. 
Due to this result $d$-majorization is suited to analyse reachable sets in the toy model (cf.~Part I of this talk)---yet as soon as one considers $n$-level quantum systems one needs a similar concept on (density) matrices.
\subsection{Generalizing $d$-Majorization to Matrices}

Classical majorization on the level of hermitian matrices uses their ``eigenvalue vector'' $\vec\lambda(\cdot)$ arranged in any order with multiplicities counted. %\\\\
%\textcolor{red}{$\prec$ nicht definiert;---der Zusammenhang zwischen $d$ und klassicher Majorisierung fehlt; eventuell Teile des folgenden Abschnitts ``The same ...'' vorziehen (nach Sec. 2.1)}\\\\
For $A,B\in\mathbb C^{n\times n}$ hermitian, $A$ is said to be majorized by $B$ if $\vec\lambda(A)\prec\vec\lambda(B)$, cf. \cite{Ando89}. The most na{\"i}ve approach to define \mbox{$D$-majorization} on matrices (with\footnote{Here and henceforth $\operatorname{diag}(x)\in\mathbb C^{n\times n}$ is the matrix which has $x\in\mathbb C^n$ on its diagonal and the remaining entries are $0$.} $D=\operatorname{diag}(d)$ for some $d\in\mathbb R_{++}^n$) would be to replace $\prec$ by $\prec_d$ and leave the rest as it is.
However
% as in the vector case
such a definition is unfeasible because it depends on the arrangement of the eigenvalues in $\vec\lambda$, due to the lack of permutation invariance of $d$ (unless $d=e^T$).
%\marginpar{Der Einschub ``as in the vector case'' ist schlecht nachvollziehbar!}

The most natural way out of this dilemma is to remember that classical majorization on matrices can be equivalently characterised via linear maps which are completely positive and trace-preserving (\textsc{cptp}) and which have the identity matrix $\operatorname{id}=\operatorname{diag}(1,\ldots,1)$ as a fixed point.
Therefore it seems utmost reasonable to generalize $d$-majorization on square matrices as follows:

\textit{Definition 1.} Given $n\in\mathbb N$ and $A,B\in\mathbb C^{n\times n}$ as well as a positive definite matrix $D\in\mathbb C^{n\times n}$ we say that $A$ is \mbox{$D$-majorized} by $B$ (denoted by $A \prec_D B$) if there exists a \textsc{cptp} map $T$ such that $T(B)=A$ and $T(D)=D$. 

Such a definition is also justified by the following: given real vectors $x,y$ and a positive vector $d\in\mathbb R_{++}^n$ one can show that $\operatorname{diag}(x)\prec_{\operatorname{diag}(d)} \operatorname{diag}(y)$ if and only if $x\prec_dy$. In other words the diagonal case reduces to $d$-majorization on vectors as expected.

Be aware that one \textit{could} define matrix $D$-majorization via positive (instead of completely positive) trace-preserving maps, and that this would make a conceptual difference -- unless $D\neq\operatorname{id}$ \cite[Thm.~7.1]{Ando89}, more on this at the end of Section \ref{sec_char_D_Maj}. However, we defined $D$-majorization via \textsc{cptp} maps because this class has a richer theory behind it and because it is the more natural choice if one comes from quantum information and control.

%Moreover Definition 1 allows for a physical interpretation of \mbox{$D$-majorization}: the identity matrix is the dynamical fixed point of any $n$-level system (with Hamiltonian $H_0\in\mathbb C^{n\times n}$) coupled to a bath of infinite temperature because the Gibbs state (i.e.~the thermodynamic equilibrium state) of such a system is
%$$
%\lim_{T\to\infty}\frac{\exp(-H_0/T)}{\operatorname{tr}(\exp(-H_0/T))}=\frac{\operatorname{id}_n}{n}=\operatorname{diag}\Big(\frac1n,\ldots,\frac1n\Big)\,.
%$$

\section{Properties of $D$-Majorization}

Using \textsc{cptp} maps in Definition 1 also allows for a physical interpretation of \mbox{$D$-majorization}: Given some $n$-level system (with Hamiltonian $H_0\in\mathbb C^{n\times n}$) coupled to a bath of some temperature $T> 0$, the Gibbs state (that is, the thermodynamic equilibrium state) of the system is given by 
$$
\rho_\text{Gibbs}^{H_0,T}:=\frac{\exp(-H_0/T)}{\operatorname{tr}(\exp(-H_0/T))}>0\,.
$$
Because every positive definite $n\times n$ matrix of unit trace is the Gibbs state of \textit{some} $n$-level system this links \mbox{$D$-majorization} to Gibbs-preserving \textsc{cptp} maps. Moreover in the high-temperature limit the above definition reduces to
$
\lim_{T\to\infty}\rho_\text{Gibbs}^{H_0,T}=\frac1n\operatorname{diag}(1,\ldots,1)\,,
$
which connects classical majorization to baths of infinite temperature.

\subsection{Characterizations of $D$-Majorization}\label{sec_char_D_Maj}

An important observation is that for any $A,B\in\mathbb C^{n\times n}$ and $D>0$ one has $A\prec_D B$ if and only if $UAU^*\prec_{UDU^*}UBU^*$ for all unitary matrices $U\in\mathbb C^{n\times n}$. Thus we can w.l.o.g.~assume that $D$ is diagonal in the standard basis.

Now if one deals with qubits, i.e.~two-dimensional systems, then \mbox{$D$-majorization} can be characterized as follows. 

{\it Proposition 2. Let $d\in\mathbb R_{++}^2$, $D=\operatorname{diag}(d)$ and $A,B\in \mathbb C^{2\times 2}$ hermitian be given. The following are equivalent.
\begin{itemize}
\item[(i)] $A\prec_D B$\smallskip
\item[(ii)] There exists a positive trace-preserving map $T$ with $T(D)=D$ and $T(B)=A$.\smallskip
\item[(iii)] $\|A-tD\|_1\leq \|B-tD\|_1$ for all $t\in\mathbb R$ with $\|\cdot\|_1$ being the trace norm.\smallskip
%\item[(v)] $\tr(A)=\tr(B)$ and $\|A-tD\|_1\leq \|B-tD\|_1$ for all $t\in\sigma(D^{-1/2}BD^{-1/2})$ where $\sigma(\cdot)$ denotes the spectrum.
\item[(iv)] $\tr(A)=\tr(B)$ and $\|A-b_iD\|_1\leq \|B-b_iD\|_1$ for $i=1,2$ as well as for the generalized fidelity
$$
\big\|\sqrt{A-b_1D}\sqrt{b_2D-A}\big\|_1\geq \big\|\sqrt{B-b_1D}\sqrt{b_2D-B}\big\|_1\,.
$$
Here $\sigma(D^{-1/2}BD^{-1/2})=\{b_1,b_2\}$ ($b_1\leq b_2$) with $\sigma(\cdot)$ being the spectrum% as well as $F(P_1,P_2):=\tr\big(\sqrt{ P_1^{1/2}P_2P_1^{1/2} }\big)$ for any $P_1,P_2\in\pos n$ denoting the generalized fidelity
.
\end{itemize}}

Of course property (iv) is the closest to the $1$-norm characterization of $\prec_d$ from Sec.~\ref{sec:d_maj_vec} and, moreover, the key to easily check (e.g., on a computer) if some hermitian matrix \mbox{$D$-majorizes} another. Unfortunately \textit{none} of these characterizations generalize to dimensions larger than $2$ because the counterexample to the Alberti-Uhlmann theorem in higher dimensions, given by \cite{HeinosaariWolf12}, pertains to our problem: Consider the hermitian matrices
\begin{equation}\label{eq:counterex_heinosaari}
A=\begin{pmatrix} 2&1&0\\1&2&-i\\0&i&2 \end{pmatrix}\quad B=\begin{pmatrix} 2&1&0\\1&2&i\\0&-i&2 \end{pmatrix}\quad D=\begin{pmatrix} 2&1&0\\1&2&1\\0&1&2 \end{pmatrix}\,.
\end{equation}
Then $\sigma(D)=\{2,2+\sqrt{2},2-\sqrt{2}\}$ so $D>0$. Obviously, $B^T=A$ and $D^T=D$ so because the transposition map is well-known to be linear, positivity- and trace-preserving one has
$
\|A-tD\|_1=\|(B-tD)^T\|_1= \|B-tD\|_1
$
for all $t\in\mathbb R$. But there exists no \textsc{cptp} map, i.e.~no $T\in Q(n)$ such that $T(B)=A$ and $T(D)=D$ as shown in \cite[Proposition 6]{HeinosaariWolf12}. For now finding simple-to-verify conditions for $\prec_D$ beyond two dimensions remains an open problem.
\subsection{Order Properties of $D$-Majorization}
As is readily verified $\prec_d$ is a preorder but it is not a partial order---the same holds for $\prec_D$ and the counterexample which shows that $\prec_d$ is not a partial order transfers to the matrix case. Moreover, now, one can characterize minimal and maximal elements in this preorder.  

{\it Theorem 3. Let $d\in\mathbb R_{++}^n$ be given and let
\begin{align*}
\mathfrak h_d&:=\lbrace X\in\mathbb C^{n\times n}\,|\,X\text{ hermitian and }\tr(X)=\unitvector^Td\rbrace\\
\mathfrak h_d^+&:=\lbrace X\in\mathbb C^{n\times n}\,|\,X\geq 0\text{ and }\tr(X)=\unitvector^Td\rbrace
\end{align*}
be the trace hyperplane induced by $d$ within the hermitian and the positive semi-definite matrices, respectively. The following statements hold.
\begin{itemize}
\item[(i)] $D=\diag d$ is the unique minimal element in $\mathfrak h_d$ with respect to $\prec_D$.\smallskip
\item[(ii)] $(\unitvector^Td) e_ke_k^T$ is maximal in $\mathfrak h_d^+$ with respect to $\prec_D$ where $k$ is chosen such that $d_k$ is minimal in $d$. It is the unique maximal element in $\mathfrak h_d^+$ with respect to $\prec_D$ if and only if $d_k$ is the unique minimal element of $d$.
\end{itemize}}

From a physical point of view this is precisely what one expects: from the state with the largest energy one can generate every other state (in an equilibrium-preserving manner) and there is no other state with this property.

\subsection{Reachable Sets \& $D$-Majorization}\label{sec_reach}
Let us finally connect our notion of $D$-majorization to the reachability questions we touched upon in the introduction. Markovian quantum control systems are generally modelled via a controlled \textsc{gksl}-equation [\cite{GKS76,Lindblad76}]:
\begin{equation}\label{eq:control_gksl}
\dot\rho(t)=-i\Big[H_0+\sum\nolimits_{j=1}^mu_j(t)H_j,\rho(t)\Big]-\gamma(t)\Gamma(\rho(t))
\end{equation}
with initial state $\rho(0)=\rho_0\in\mathbb C^{n\times n}$, control Hamiltonians $H_1,\ldots,H_m$, and control amplitudes $u_1,\ldots,u_m,\gamma$. Here $\Gamma(\rho):=\sum_{j\in I}(\frac12(V_j^*V_j\rho+\rho V_j^*V_j)-V_j\rho V_j^*)$ describes the dissipative effect on the system by means of the matrices $(V_j)_{j\in I}\subset\mathbb C^{n\times n}$ which in principle can be arbitrary.

Now given any $n$-level system described by a hermitian matrix $H_S\in\mathbb C^{n\times n}$ with spectral decomposition $\sum_{j=1}^n E_j|g_j\rangle\langle g_j|$, $E_1\leq\ldots\leq E_n$ and a bath of some temperature $T>0$ the coupling of the system to said bath can be modelled by \eqref{eq:control_gksl} if the generators of the dissipation $(V_j)_{j\in I}$ are chosen to be the modified ladder operators
\begin{equation}\label{eq:sigma_plus_minus}
\begin{split}
\sigma_+^d&=:\sum\nolimits_{j=1}^{n-1} \sqrt{\frac{j(n-j)e^{-E_j/T}}{e^{-E_j/T}+e^{-E_{j+1}/T}}  }\,|g_j\rangle\langle g_{j+1}|\\
\sigma_-^d&=:\sum\nolimits_{j=1}^{n-1} \sqrt{\frac{j(n-j)e^{-E_{j+1}/T}}{e^{-E_j/T}+e^{-E_{j+1}/T}}  }\,|g_{j+1}\rangle\langle g_{j}|\,.
\end{split}
\end{equation}

In order to analyze the reachable set of \eqref{eq:control_gksl} with $H_0=H_S$ and dissipation generators $\sigma_+^d,\sigma_-^d$ we (as in Section \ref{sec:d_maj_vec}) define the set of all matrices which are \mbox{$D$-majorized} by some state $\rho$ or a collection of states $S\subseteq\mathbb C^{n\times n}$: 
\begin{align*}
M_D:\mathcal P(\mathbb C^{n\times n})&\to\mathcal P(\mathbb C^{n\times n})\\
S&\mapsto \bigcup\nolimits_{\rho\in S}\lbrace X\in\mathbb C^{n\times n}\,|\,X\prec_D \rho\rbrace
\end{align*}
with $\mathcal P$ being the power set and $M_D(X):=M_D(\{X\})$ for all $X\in\mathbb C^{n\times n}$. This operator is used to upper bound the reachable set of the ``toy model'' $\Lambda_d$ (cf.~Part I)\footnote{
More precisely we proved that $\mathfrak{reach}_{\Lambda_d}(x_0)\subseteq (M_e\circ M_d)(x_0) $ for any initial state $x_0$ and $d\in\mathbb R_{++}^n$ corresponding to a spin system, i.e.~$d=(\alpha^{j-1})_{j=1}^n$ for some $\alpha\in(0,1)$.
\label{footnote:4}} and is expected to do so in the matrix case, as well. Important properties of $M_D$ are:
\begin{itemize}
\item[(i)] $M_D(X)$ is convex for all $X\in\mathbb C^{n\times n}$.\smallskip
\item[(ii)] If $P\subset\mathbb C^{n\times n}$ is compact, then $ M_D(P)$ is compact.\smallskip
\item[(iii)] If $P$ is a collection of quantum states then $M_D(P)$ is star-shaped with respect to the Gibbs state $\frac{D}{\operatorname{tr}(D)}$.
\item[(iv)] When restricting $M_D$ to the compact subsets of $\mathbb C^{n\times n}$ then $M_D$ is non-expansive (so in particular continuous) with respect to the Hausdorff metric.
\end{itemize}

The last property formulates that for a system %described by a Hamiltonian $H_0$ 
in the state $\rho$ which is coupled to a bath of temperature $T\geq 0$, ``small'' changes in $\rho$ cannot change the set of \mbox{$D$-majorized} states ``too much''. 
%This sparks a physically interesting and relevant question: is the set of \mbox{$D$-majorized} states also robust to ``small'' changes in temperature of the bath? In other words is the map
%\begin{align*}
%M^P:(\mathfrak{pos}(n),\|\cdot\|_1)&\to\mathcal P(\mathbb C^{n\times n})\\
% D&\mapsto M_D(P)
%\end{align*}
%(for arbitrary compact $P\subset\mathbb C^{n\times n}$) continuous in the Hausdorff metric? 

Coming back to footnote \ref{footnote:4}, the crucial step in the proof is to identify an extreme point of $M_d(x_0)$ which is maximal w.r.t.~classical majorization. While an extreme point analysis of the set of matrices $M_D(\rho_0)$ is way more difficult---as the convex polytope techniques from the vector case break down---the idea of a maximal extreme point might be equally useful in analyzing general open quantum control problems in the future.

\section{Connection to Thermo-Majorization}

Over the last few years, sparked by \cite{Brandao15,Horodecki13} and others [\cite{Gour15,Lostaglio18,Sagawa19}] thermo-majorization has been a widely discussed and studied
topic in quantum physics and in particular quantum thermodynamics. In the abelian case thermo-majorization, on a mathematical level, is described by vector $d$-majorization which begs the question of how to define thermo-majorization for general quantum states.

Indeed \cite{Faist17} have shown that it makes a conceptual difference whether one defines thermo-majorization on non-diagonal states via Gibbs-preserving maps (i.e.~\textsc{cptp} maps having the Gibbs state $D>0$ as a fixed point, cf.~Definition 1) or if one restricts to the smaller class of thermal operations. The latter, given some Hamiltonian of the system $H_S$ and a fixed bath temperature $T\geq 0$, are defined as follows, cf.~also \cite{Lostaglio19}:

\textit{Definition 4.} A linear map $\Phi:\mathbb C^{n\times n}\to\mathbb C^{n\times n}$ is a thermal operation w.r.t.~$H_S$ if there exist $m\in\mathbb N$, $H_R\in\mathbb C^{m\times m}$ hermitian, and $U\in\mathbb C^{mn\times mn}$ unitary such that
\begin{equation}\label{eq:energy_conserv}
[U,H_S\otimes\operatorname{id}_R+\operatorname{id}_S\otimes H_R]=0
\end{equation}
and
$$
\Phi(\rho)=\operatorname{tr}_R(U(\rho\otimes\rho_{\text{Gibbs}}^{H_R,T})U^*)
$$
for all $\rho\in\mathbb C^{n\times n}$ (or equivalently for all quantum states $\rho$). We denote the collection of all thermal operations by $\mathsf{TO}(H_S,T)$.

Thermal operations are the free operations of the resource theory of quantum thermodynamics as those encompass the dynamics which preserve the Gibbs-state and which satisfy \eqref{eq:energy_conserv} (conserve the global energy, that is, the energy of the larger system \mbox{$H_{SR}=H_S\otimes\operatorname{id}_R+\operatorname{id}_S\otimes H_R$}).
One readily verifies that $\mathsf{TO}(H_S,T)$ forms a path-connected semigroup with identity and -- although $\mathsf{TO}(H_S,T)$ in general is not closed -- its closure even is convex and compact. 

In the vector case the state transitions possible with thermal operations are the same as with general Gibbs-preserving maps described by $d$-stochastic matrices. However in the operator case there is a discrepancy between the two coming from the fact that there exist Gibbs-preserving maps which generate coherent superpositions of energy levels, whereas no thermal operation is capable of doing such a thing. In fact for all $H_S\in\mathbb C^{n\times n}$, $T\geq 0$ one finds the inclusions
\begin{equation}\label{eq:TO_incl}
\mathsf{TO}(H_S,T)\subseteq\mathsf{EnTO}(H_S,T)\subsetneq Q_{e^{-H_S/T}}(n)
\end{equation}
where $Q_{e^{-H_S/T}}(n)$ is the collection of all \textsc{cptp} maps which have $e^{-H_S/T}$ -- and thus $\rho_\text{Gibbs}^{H_S,T}$ -- as a fixed point, and 
$$
\mathsf{EnTO}(H_S,T):=\{\Phi\in Q_{e^{-H_S/T}}(n)\,:\,[\Phi,\operatorname{ad}_{H_S}]=0\}
$$
are the enhanced thermal operations (also called ``covariant Gibbs-preserving maps'').
This is an important observation as the covariance property $[\Phi,\operatorname{ad}_{H_S}]=0$ forces that the diagonal and the off-diagonal action of any channel are strictly separated, assuming $H_S$ has non-degenerate spectrum.
Note that this insight is of importance to us because the solution to the uncontrolled master equation \eqref{eq:control_gksl} (i.e.~$H_1=\ldots=H_m=0$, $\gamma\equiv 1$) with dissipation generators $\sigma_+^d,\sigma_-^d$ from \eqref{eq:sigma_plus_minus} lives in $\mathsf{EnTO}$ at all times.

Be aware that in \eqref{eq:TO_incl}, if the thermal operations are replaced by their closure then the first set inclusion is an equality if $n=2$ and becomes a strict inclusion for $n\geq 3$ as shown by \cite{Ding21}. Even worse this discrepancy between $\mathsf{TO}$ and $\mathsf{EnTO}$ remains when looking at the \textit{action} of the respective sets on certain states; more precisely, there exist quantum states $\rho,\omega\in\mathbb C^{3\times 3}$ and an enhanced thermal operation $\Phi$ such that $\Phi(\rho)=\omega$ but no element in $\mathsf{TO}$ or its closure can map $\rho$ to $\omega$.

%On the other hand there are some properties one, mathematically and physically, expects from a notion of majorization connected to a temperature $T\in[0,\infty]$ which are unclear or even amiss if one is restricted to thermal operations:
%\begin{itemize}
%\item[$\bullet$] Do the thermal operations form a semigroup, i.e.~is the composition of two thermal operations again a thermal operation? Physical intuition suggests that this should hold but mathematically this seems unclear. This semigroup property is required to guarantee that thermo-majorization is a preorder, which is a justified expectation: if one can generate state $A$ from state $B$ and state $B$ from state $C$ then one should be able to generate state $A$ from state $C$ (all via ``free operations'').\smallskip
%\item[$\bullet$] There is no maximal state anymore (in the sense of Thm.~3). The state of %(unique)
% largest energy loses this property as shown by the same example which demonstrates the conceptual difference between Gibbs-preserving maps and thermal operations (\cite{Faist17}). Moreover, because the state with largest energy is uniquely maximal in $\prec_D$ there cannot be a replacement if one is restricted to a smaller class of \textsc{cptp} 
% maps.
%\end{itemize}
This observation is particularly important for the field of quantum control as there one usually wonders which state transitions can be realized under a given control scenario. Thus beyond qubits it makes a conceptional difference which of the sets in \eqref{eq:TO_incl} one uses to model a given thermodynamic control problem.
Moreover, quantum control problems usually come in the framework of quantum-dynamical semigroups so one in addition needs to identify those quantum maps from a
certain set (usually carrying the structure of a semigroup) which can be written as the solution to a controlled master equation of Gorini-Kossakowski-Sudarshan-Lindblad type [\cite{GKS76,Lindblad76}], that is, to identify those channels which are time-dependent Markovian. In other words from a Lie-theoretical perspective
one wants to determine the Lie wedge of the respective semigroup in order to characterize the desired quantum channels as solutions of suitable
  (bi)linear master equations.

The question of Markovian state transitions in thermodynamics has only been tackled recently by \cite{LosKor21} for the set of enhanced thermal operations and the simpler case of diagonal states -- recall that in this realm the problem reduces to vector-$d$ majorization. They were able to fully characterize which state transitions are possible under Markovian thermal processes (i.e.~maps from $\mathsf{EnTO}$ which are solutions of a time-dependent \textsc{gksl}-equation) in the classical realm, and they even gave algorithms to check for a Markovian path from a given initial to a given final state.
%\marginpar{Hier ist ein ``is'' zuviel.}
While incredibly important, their work of course is but a first step in this direction and the ultimate goal will be to extend their results and concepts to general thermodynamic quantum control systems.

\bibliography{../../../../control21vJan20}            % bib file to produce the bibliography

\begin{thebibliography}{19}
\providecommand{\natexlab}[1]{#1}
\providecommand{\url}[1]{\texttt{#1}}
\providecommand{\urlprefix}{URL }
\expandafter\ifx\csname urlstyle\endcsname\relax
  \providecommand{\doi}[1]{doi:\discretionary{}{}{}#1}\else
  \providecommand{\doi}{doi:\discretionary{}{}{}\begingroup
  \urlstyle{rm}\Url}\fi

\bibitem[{Ando(1989)}]{Ando89}
Ando, T. (1989).
\newblock {Majorization, Doubly Stochastic Matrices, and Comparison of
  Eigenvalues}.
\newblock \emph{Lin. Alg. Appl.}, 118, 163--248.

\bibitem[{Brand{\~a}o et~al.(2015)Brand{\~a}o, Horodecki, Ng, Oppenheim, and
  Wehner}]{Brandao15}
Brand{\~a}o, F., Horodecki, M., Ng, N., Oppenheim, J., and Wehner, S. (2015).
\newblock {The Second Laws of Quantum Thermodynamics}.
\newblock \emph{Proc. Natl. Acad. Sci. U.S.A.}, 112, 3275--3279.

\bibitem[{Ding et~al.(2021)Ding, Ding, and Hu}]{Ding21}
Ding, Y., Ding, F., and Hu, X. (2021).
\newblock {Exploring the Gap Between Thermal Operations and Enhanced Thermal
  Operations}.
\newblock \emph{Phys. Rev. A}, 103, 052214.

\bibitem[{Dirr et~al.(2019)Dirr, vom Ende, and Schulte-Herbr{\"u}ggen}]{CDC19}
Dirr, G., vom Ende, F., and Schulte-Herbr{\"u}ggen, T. (2019).
\newblock {Reachable Sets from Toy Models to Controlled Markovian Quantum
  Systems}.
\newblock \emph{Proc. IEEE Conf. Decision Control (IEEE-CDC)}, 58, 2322.

\bibitem[{Faist et~al.(2015)Faist, Oppenheim, and Renner}]{Faist17}
Faist, P., Oppenheim, J., and Renner, R. (2015).
\newblock {Gibbs--Preserving Maps Outperform Thermal Operations in the Quantum
  Regime}.
\newblock \emph{New J. Phys.}, 17, 1--4.

\bibitem[{Gorini et~al.(1976)Gorini, Kossakowski, and Sudarshan}]{GKS76}
Gorini, V., Kossakowski, A., and Sudarshan, E. (1976).
\newblock {Completely Positive Dynamical Semigroups of $N$-Level Systems}.
\newblock \emph{J. Math. Phys.}, 17, 821--825.

\bibitem[{Gour et~al.(2015)Gour, M{\"u}ller, Narasimhachar, Spekkens, and
  Halpern}]{Gour15}
Gour, G., M{\"u}ller, M., Narasimhachar, V., Spekkens, R., and Halpern, N.
  (2015).
\newblock {The Resource Theory of Informational Nonequilibrium in
  Thermodynamics}.
\newblock \emph{Phys. Rep.}, 583, 1--58.

\bibitem[{Heinosaari et~al.(2012)Heinosaari, Jivulescu, Reeb, and
  Wolf}]{HeinosaariWolf12}
Heinosaari, T., Jivulescu, M., Reeb, D., and Wolf, M. (2012).
\newblock {Extending Quantum Operations}.
\newblock \emph{J. Math. Phys.}, 53, 102208.

\bibitem[{Horodecki and Oppenheim(2013)}]{Horodecki13}
Horodecki, M. and Oppenheim, J. (2013).
\newblock {Fundamental Limitations for Quantum and Nanoscale Thermodynamics}.
\newblock \emph{Nat. Commun.}, 4, 2059.

\bibitem[{Joe(1990)}]{Joe90}
Joe, H. (1990).
\newblock {Majorization and Divergence}.
\newblock \emph{J. Math. Anal. Appl.}, 148, 287--305.

\bibitem[{Lindblad(1976)}]{Lindblad76}
Lindblad, G. (1976).
\newblock {On the Generators of Quantum Dynamical Semigroups}.
\newblock \emph{Commun. Math. Phys.}, 48, 119--130.

\bibitem[{Lostaglio(2019)}]{Lostaglio19}
Lostaglio, M. (2019).
\newblock {An Introductory Review of the Resource Theory Approach to
  Thermodynamics}.
\newblock \emph{Rep. Prog. Phys.}, 82, 114001.

\bibitem[{Lostaglio et~al.(2018)Lostaglio, Alhambra, and Perry}]{Lostaglio18}
Lostaglio, M., Alhambra, {\'A}., and Perry, C. (2018).
\newblock {Elementary Thermal Operations}.
\newblock \emph{Quantum}, 2, 1--52.

\bibitem[{Lostaglio and Korzekwa(2021)}]{LosKor21}
Lostaglio, M. and Korzekwa, K. (2021).
\newblock {Continuous Thermomajorization and a Complete Set of Laws for
  Markovian Thermal Processes}.

\bibitem[{Marshall et~al.(2011)Marshall, Olkin, and Arnold}]{MarshallOlkin}
Marshall, A., Olkin, I., and Arnold, B. (2011).
\newblock \emph{{Inequalities: Theory of Majorization and Its Applications}}.
\newblock Springer, New York, 2 edition.

\bibitem[{Ruch et~al.(1978)Ruch, Schranner, and Seligman}]{Ruch78}
Ruch, E., Schranner, R., and Seligman, T. (1978).
\newblock {The Mixing Distance}.
\newblock \emph{J. Chem. Phys.}, 69, 386--392.

\bibitem[{Sagawa et~al.(2021)Sagawa, Faist, Kato, Matsumoto, Nagaoka, and
  Brand{\~a}o}]{Sagawa19}
Sagawa, T., Faist, P., Kato, K., Matsumoto, K., Nagaoka, H., and Brand{\~a}o,
  F. (2021).
\newblock {Asymptotic Reversibility of Thermal Operations for Interacting
  Quantum Spin Systems via Generalized Quantum Stein’s Lemma}.
\newblock \emph{J. Phys. A}, 54, 495303.

\bibitem[{Veinott(1971)}]{Veinott71}
Veinott, A. (1971).
\newblock {Least $d$-Majorized Network Flows with Inventory and Statistical
  Applications}.
\newblock \emph{Manag. Sci.}, 17, 547--567.

\bibitem[{vom Ende and Dirr(2019)}]{vomEnde19polytope}
vom Ende, F. and Dirr, G. (2019).
\newblock {The $d$-Majorization Polytope}.

\end{thebibliography}
                                                     % with bibtex (preferred)
                                                   
%\begin{thebibliography}{xx}  % you can also add the bibliography by hand

%B.C. Able.
%\newblock Nucleic acid content of microscope.
%\newblock \emph{Nature}, 135:\penalty0 7--9, 1956.

%\bibitem[Able et~al.(1954)Able, Tagg, and Rush]{AbTaRu:54}
%B.C. Able, R.A. Tagg, and M.~Rush.
%\newblock Enzyme-catalyzed cellular transanimations.
%\newblock In A.F. Round, editor, \emph{Advances in Enzymology}, volume~2, pages
%  125--247. Academic Press, New York, 3rd edition, 1954.

%\bibitem[Keohane(1958)]{Keo:58}
%\newblock \emph{Power and Interdependence: World Politics in Transitions}.
%\newblock Little, Brown \& Co., Boston, 1958.

%\bibitem[Powers(1985)]{Pow:85}
%\newblock Is there a way out?
%\newblock \emph{Harpers}, pages 35--47, June 1985.

%\bibitem[Soukhanov(1992)]{Heritage:92}
%\newblock \emph{{The American Heritage. Dictionary of the American Language}}.
%\newblock Houghton Mifflin Company, 1992.

%\end{thebibliography}

%\appendix
%\section{A summary of Latin grammar}    % Each appendix must have a short title.
%\section{Some Latin vocabulary}              % Sections and subsections are supported  
                                                                         % in the appendices.
\end{document}